\documentclass[11pt,twoside]{article}
\usepackage{asp2004}
\usepackage{psfig}
\usepackage{epsf}
\usepackage{graphics}
\usepackage{lscape}
\def\deg{\ifmmode ^{\rm o} \else $^{\rm o}$\fi}

\def\arcsec{\ifmmode'' \else $''$\fi}
\def\arcmin{\ifmmode ' \else $'$\fi}
\def\arcsecpoint{\ifmmode ``\!. \else $''\!.$\fi}
\def\arcminpoint{\ifmmode `\!. \else $'\!.$\fi}
\def\kms{\ifmmode {\rm km\ s}^{-1} \else km s$^{-1}$\fi}
\def\Msun{\ifmmode M_{\odot} \else $M_{\odot}$\fi}
\def\Lsun{\ifmmode L_{\odot} \else $L_{\odot}$\fi}
\def\qo{\ifmmode q_{\rm o} \else $q_{\rm o}$\fi}
\def\Ho{\ifmmode H_{\rm o} \else $H_{\rm o}$\fi}
\def\ho{\ifmmode h_{\rm o} \else $h_{\rm o}$\fi}

\def\vFWHM{\ifmmode v_{\mbox{\tiny FWHM}} \else
            $v_{\mbox{\tiny FWHM}}$\fi}
\def\CCF{\ifmmode F_{\it CCF} \else $F_{\it CCF}$\fi}
\def\ACF{\ifmmode F_{\it ACF} \else $F_{\it ACF}$\fi}
\def\Halpha{\ifmmode {\rm H}\alpha \else H$\alpha$\fi}
\def\Hbeta{\ifmmode {\rm H}\beta \else H$\beta$\fi}
\def\Hgamma{\ifmmode {\rm H}\gamma \else H$\gamma$\fi}
\def\Hdelta{\ifmmode {\rm H}\delta \else H$\delta$\fi}
\def\Lya{\ifmmode {\rm Ly}\alpha \else Ly$\alpha$\fi}
\def\Lyb{\ifmmode {\rm Ly}\beta \else Ly$\beta$\fi}

\def\ciii{\ifmmode {\rm C}\,{\sc iii} \else C\,{\sc iii}\fi}
\def\civ{\ifmmode {\rm C}\,{\sc iv} \else C\,{\sc iv}\fi}

\def\o5007{[O\,{\sc iii}]\,$\lambda5007$}

\begin{document}

\title{Fast Optical Photometry of Galaxies: Observations of Short-Lived Flare Events}
\author{B.E.\ Zhilyaev, Ya.O.\ Romanyuk, I.A.\ Verlyuk,
O.A. Svyatogorov, and M.~I.~Petrov}

\affil{Main Astronomical Observatory, NAS of Ukraine,\\ 27
Akademika Zabolotnoho Str., 03680 Kiev, Ukraine}

\author{M.N.\ Lovkaya}
\affil{Crimean Astrophysical Observatory, Nauchny, 98409 Crimea,
Ukraine}

\begin{abstract}
We have monitored two bright galaxies, M\,85 and NGC 7331, on
timescales as short as  0.01\,s. In the optical, we discovered an unusual
burst coincident with the galaxy M\,85. We registered a sudden
onset with a characteristic time of less than 10\,ms with
subsequent quasi-exponential decay within approximately 1\,s and
an amplitude of 2.5\,mag in the $V$ band. In the course of
high-speed monitoring with two Crimean telescopes operated
synchronously, in both independent instruments we have registered
one coincident event occurring in NGC\,7331 with a duration of
$\sim 0.6$\,s. The amplitudes range from $\sim 3$\,mag to $\sim 0.3$\,mag
in the $U$ and $I$ bands, respectively. Merging of an
intermediate-mass black hole with a small black hole or normal star seems to be
the most plausible mechanism responsible for short bursts. Our
observations support the hypothesis concerning the existence of
intermediate-mass black holes in the centers of galaxies and in dense
globular clusters.
\end{abstract}

\section{Introduction}
We find evidence for variability of some galactic nuclei in
the $U\!BV\!RI$ bands on a timescale of tenths of a second. Byrne \&
Wayman (1975) first carried out a search for optical flashes in
the direction of the center of the Galaxy in the early 1970s.
This experiment was the first attempt to find an outburst of
radiation that  may accompany such energetic events as pulses of
gravitational waves (GW). Recent studies of coincident events in the
gravitational wave detectors Explorer and Nautilus (Astone et al.\
2002) give an observed rate around one event per day. Such events
correspond to an isotropic conversion of $0.004\,\Msun$ into GW
energy for sources located in our Galaxy. This gives us some hope
of detecting optical flares from the cores of other galaxies.  To detect
flares with confidence, it is desirable to carry out a high-speed
monitoring by two or more remote telescopes synchronously. We
present some preliminary results for two objects, the galaxies M\,85
(S0, $B = 10.0$\,mag) and NGC 7331 (Sb, $B = 10.3$\,mag). Both of these show
very bright inner regions suitable for high-speed monitoring in
the optical on timescales as short as $0.01\,$s with adequate
signal-to-noise ratios.

\section{Observations and Data Analysis}

Observations were made in the $U\!BV\!RI$ bands with the 2-m RCC
telescope at Peak Terskol and the Crimean 1.25-m and 50-inch
telescopes equipped with high-speed photon-counting photometers.
For the majority of these observations, these telescopes were
operated synchronously. The nuclei of galaxies were observed with
large focal-plane diaphragms (from 26\arcsec\ to 50\arcsec). Second
channels of the photometers were used for monitoring comparison
stars. In all, we obtained 98 minutes of photometric data sampled
at 100 Hz.

\subsection{Flare Identification}

Figure 1 shows an unusual burst in the optical registered while
observing the core of the galaxy M\,85 during a 1 hour-long run. The
burst consists of a sudden onset with a characteristic time of
less than $10\,$ms, with the subsequent quasi-exponential decay
within approximately $1$\,s. Its amplitude amounts to 2.5\,mag in
the $V$ band. No bursts were found in the control star that was observed
simultaneously. The maximum flux is equivalent to the flux from a
$V \approx 9.5$\,mag star at the zenith. According to the data given by
Allen (1973), the expected mean rate at the zenith for meteors
brighter than $V \leq 9.5$\,mag in a 39\arcsec\ field (focal-plane
diaphragm) is $\sim 4.8\times10^{-5}$\,hr$^{-1}$. Therefore, it is
reasonable to assume that a faint meteor did not cause the flare
event in M\,85. However, we cannot completely rule out that such an
event may be produced by some other unknown phenomenon, e.g., by a
cosmic-ray event.

\begin{figure}[!ht]
\centerline{\psfig{figure=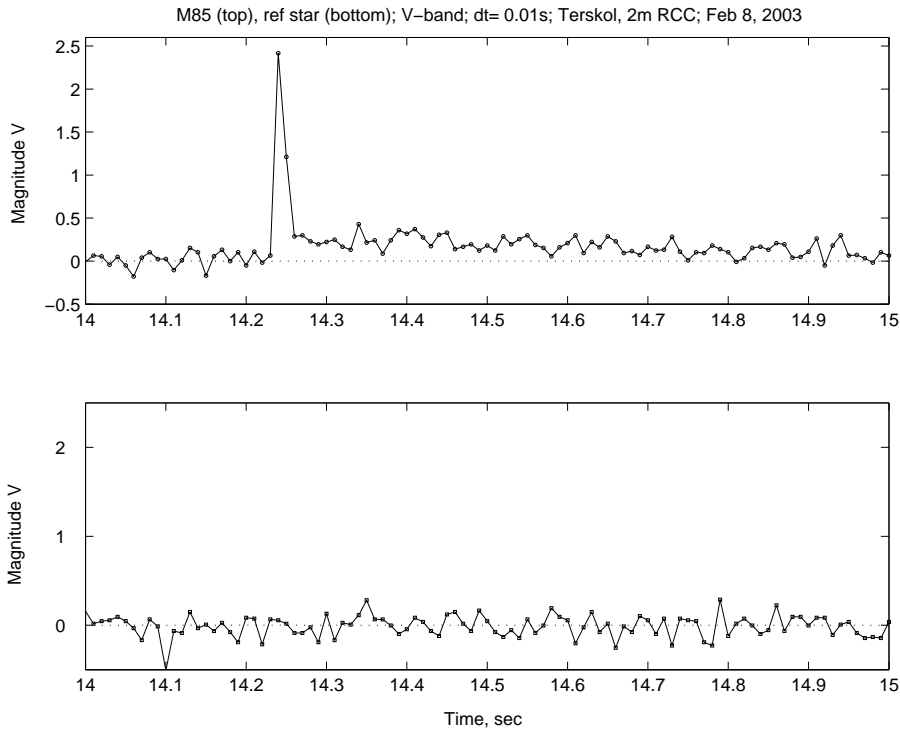,width=120truemm,angle=0,clip=}}
\caption{Fragments of $V$-band light curves the galaxy M\,85 (top)
and a reference star (bottom) as seen simultaneously by the
Terskol 2-m telescope on 2003 February 8, 02:05:22 UT (max).}
\end{figure}
During the high-speed monitoring with the two Crimean telescopes
operating synchronously, we also found one coincident event
occurring in NGC 7331 during a 38 minute-long run. This event
corresponds to a short burst with duration of $\sim 0.6$\,s (Fig.\
2). The amplitudes range from $\sim 3$\,mag to $\sim 0.3$\,mag
in the $U$ and $I$ bands, respectively.

\begin{figure}
\vskip2mm
\centerline{{\psfig{figure=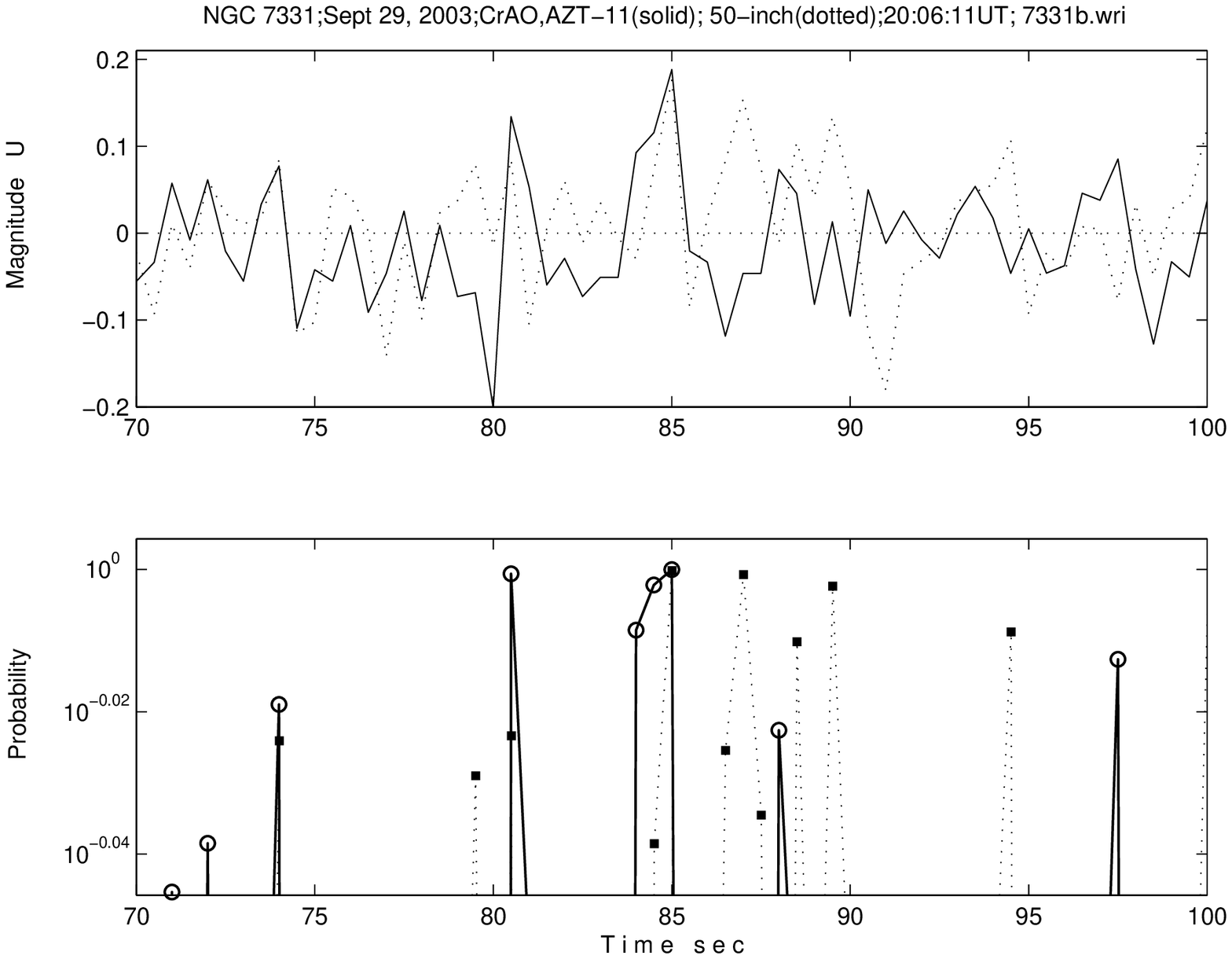,width=60truemm,angle=0,clip=}}
\hskip2mm
{\psfig{figure=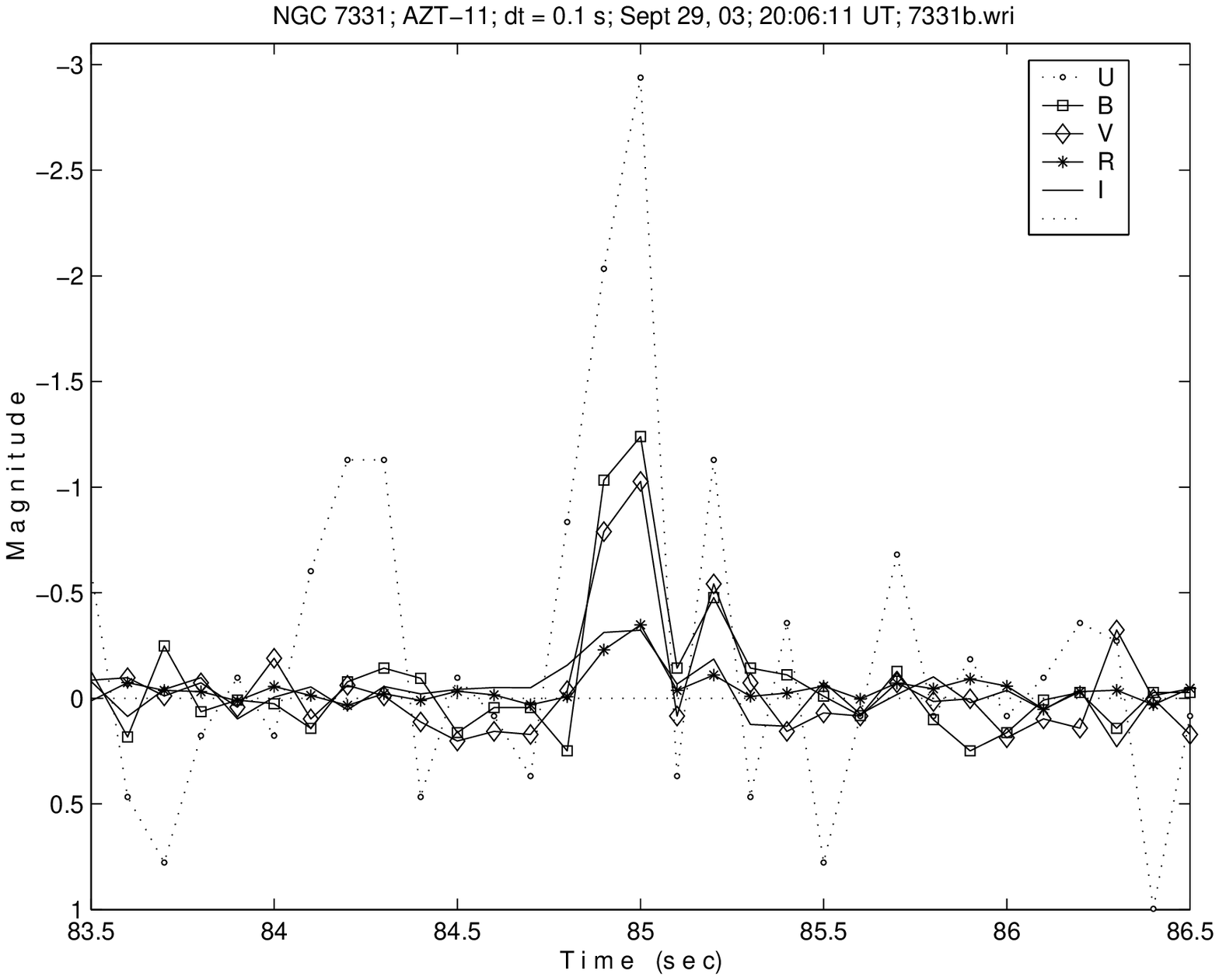,width=60truemm,angle=0,clip=}}}
\caption{Left side: The light curves of NGC 7331 from the Crimean
AZT-11 (solid) and 50-inch (dotted) telescopes (upper panel) with
a sampling time $0.5\, s$ and their cumulative probability
functions (lower panel) on September 29, 2003, 20:06:11 UT. Right
side: Fragments of NGC 7331 light curves in the UBVRI bands with a
sampling time $0.1\, s$.} \vskip2mm
\end{figure}


The detection of transient variations is based on a search for
coincidence between the data of different telescopes operating
synchronously. Let us convert the light curve using the cumulative
Poisson distribution with the mean and current readings $n_{0}$
and $n$, respectively,
\begin{equation}
  P(n,n_{0})= \sum \limits_{k=0}^{n}\frac{n_{0}^{k}}{k!}e^{-n_{0}}
\end{equation}
which is the probability of obtaining the observational result
$n$. If a coincidence between two telescopes is due to a real
burst, we expect the individual probabilities to be very close to
one. The quantity $SL=1-P(n,n_{0})$ is the significance level to
obtain the observational result $n$ in the light curve. To calculate
the joint significance for $m$ synchronously operating telescopes,
we may write
\begin{equation}
 SL_{m}= -2\sum \limits_{i=1}^{m}\ln(SL_{i})
\end{equation}
As noted by Fisher (1970), the statistic $SL_{m}$ follows the
$\chi^{2}_{2m}$ distribution with $2m$ degrees of freedom. Further
discussion of the problem can be found in our paper (Zhilyaev et
al.\ 2003).

Figure 2 shows that the coincidence technique may lead to a
substantial gain in detection of small-scale variability. The
flare event at 85 sec is defined by the individual confidence
probability of 99.999497 and 99.942578 percent, respectively. Its
joint confidence probability, which was defined in equation (2),
goes up to 99.999994 percent with two telescopes operating
synchronously. The frequency of occurrence of such an event is
equal to 0.000275 for a record of 38 minutes and sampling frequency of
2\,Hz. It means that one should perform on the average
3636 similar tests of 38 min duration each, that is, 96 days in
total, to obtain the same result on account of random fluctuations. Thus
we argue that the flare event is a real phenomenon.

\section{Discussion and Conclusions}

There is a growing evidence for the existence of intermediate-mass
black holes in the mass range $\sim 10^{2}$--$10^{4}\,\Msun$,
especially in dense globular clusters (Miller 2003). These are
undergoing frequent coalescence with stellar-mass black holes as
well as normal stars. Stars whose orbits cross the event horizon
or the tidal disruption radius of a black hole are destroyed
before they complete the orbit (Alexander et al.\ 2003).
The Keplerian
orbital period for a black hole of mass $M$ in a vicinity of
the event horizon is $P \approx 10^{-5}\,(M/\Msun)$\,s. Thus,
orbital inspiral into a supermassive black hole in a galactic
center may be a strong source of gravitational radiation (and
optical emission as well) on timescales of 10--$10^{5}$\,s.
In the case of intermediate-mass black holes, this timescale is in the
range 1--100\,ms.

We adopt the luminosity of M\,85  to be
$3\times 10^{43}\,$erg s$^{-1}$ (a typical value).
Then the M\,85 flare event has energy
$1.3 \times 10^{42}\,$erg s$^{-1}$ in the $V$ band.
We may estimate  the mass
of a compact object to be $\sim 1000\, \Msun$, if the duration of the
M\,85 flare event is defined by the orbital inspiral into a black
hole. The NGC 7331 flare event has a peak power value
$\sim 3.3 \times 10^{42}\,$erg s$^{-1}$ in the $B$ band,
close to that of the
M\,85 flare event. Merging of intermediate-mass black holes with small
black holes
or normal stars in galactic nuclei and globular clusters seems to be
the most plausible mechanism for short bursts. As shown by Misner
et al.\ (1973), during the merger the output in gravitation waves
may reach $10^{51}$--$10^{56}\,$erg s$^{-1}$; in
electromagnetic waves, the luminosity may be many orders of
magnitude less.

To summarize, our observations support the hypothesis that
intermediate-mass black holes exist in the centers of
galaxies and dense globular clusters.


%
\end{document}